\documentclass[aps,prx,twocolumn,groupedaddress,10pt]{revtex4-1}
\usepackage{amsmath}
\usepackage{graphicx}
\usepackage{color}

\newcommand{\bk}{{\bf k}}

\newcommand{\bp}{{\bf p}}

\newcommand{\beq}{\begin{equation}}
\newcommand{\beqn}{\begin{eqnarray}}
\newcommand{\eeq}{\end{equation}}
\newcommand{\eeqn}{\end{eqnarray}}

\begin{document}

\title{Model Prediction of Self-Rotating Excitons\\ in Two-Dimensional Transition-Metal Dichalcogenides}

\author{Maxim Trushin$^{1,2}$}
\author{Mark Oliver Goerbig$^3$}
\author{Wolfgang Belzig$^2$}
\affiliation{$^1$Centre for Advanced 2D Materials, National University of Singapore, 6 Science Drive 2, 117546, Singapore}
\affiliation{$^2$Department of Physics, University of Konstanz, D-78457 Konstanz, Germany}
\affiliation{$^3$Laboratoire de Physique des Solides, Univ. Paris-Sud, Universit\'e Paris-Saclay, CNRS UMR 8502, F-91405 Orsay, France}

\date{\today}

\begin{abstract}
Using the quasiclassical concept of Berry curvature we demonstrate
that a Dirac exciton---a pair of Dirac quasiparticles bound by Coulomb interactions---inevitably possesses
an intrinsic angular momentum making the exciton effectively self-rotating.
The model is applied to excitons in two-dimensional transition metal dichalcogenides,
in which the charge carriers are known to be described by a Dirac-like Hamiltonian.
We show that the topological self-rotation strongly modifies the exciton spectrum and, as a consequence,
resolves the puzzle of the overestimated two-dimensional
polarizability employed to fit earlier spectroscopic measurements.  
\end{abstract}


\maketitle

\begin{figure}
\includegraphics[width=\columnwidth]{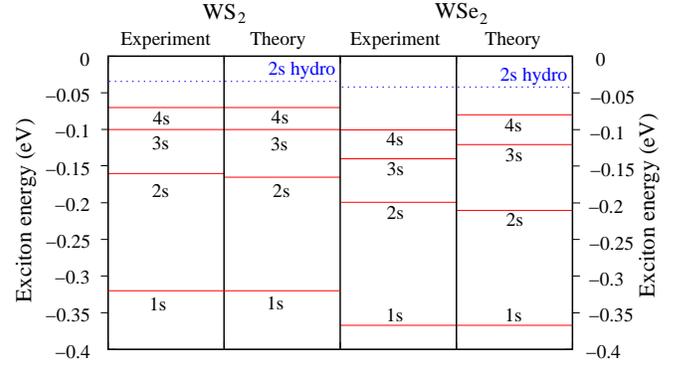}
\caption{\label{fig1} Excitonic spectrum: measurements and theory.
The experimental data have been taken from Refs. \cite{PRL2014chernikov} and \cite{PRL2014he} for WS$_2$ and WSe$_2$, respectively.
The theoretical spectra have been calculated quasiclassically from Eq.~(\ref{Coulomb2}) taking into account nonlocal screening
and self-rotation. Our theory employs reduced exciton masses $\mu=0.16 m_0$ for WS$_2$ and $\mu=0.17 m_0$ for WSe$_2$ 
(here, $m_0$ is the free electron mass)
computed {\it ab initio} in Ref.~\cite{PRB2013berkelbach}.
We assume that the screening is due to the polarizability of the 2D layer $\chi=r_0/(2\pi)$
with $r_0$ being the only fitting parameter. In contrast to the conventional model \cite{PRL2014chernikov} strongly overestimating $r_0$,
we found that $r_0=4.65$ nm for WS$_2$ and $r_0=3.84$ nm for WSe$_2$,
in agreement with the {\it ab initio} calculations \cite{PRB2013berkelbach} suggesting $r_0\sim 4$ nm in such materials.
Dotted lines show the first excited states ($2s$ states) within the standard hydrogenic model (\ref{spectrum0}) 
to emphasize the qualitative difference with our findings.}
\end{figure}

\begin{figure}
\includegraphics[width=0.495\columnwidth]{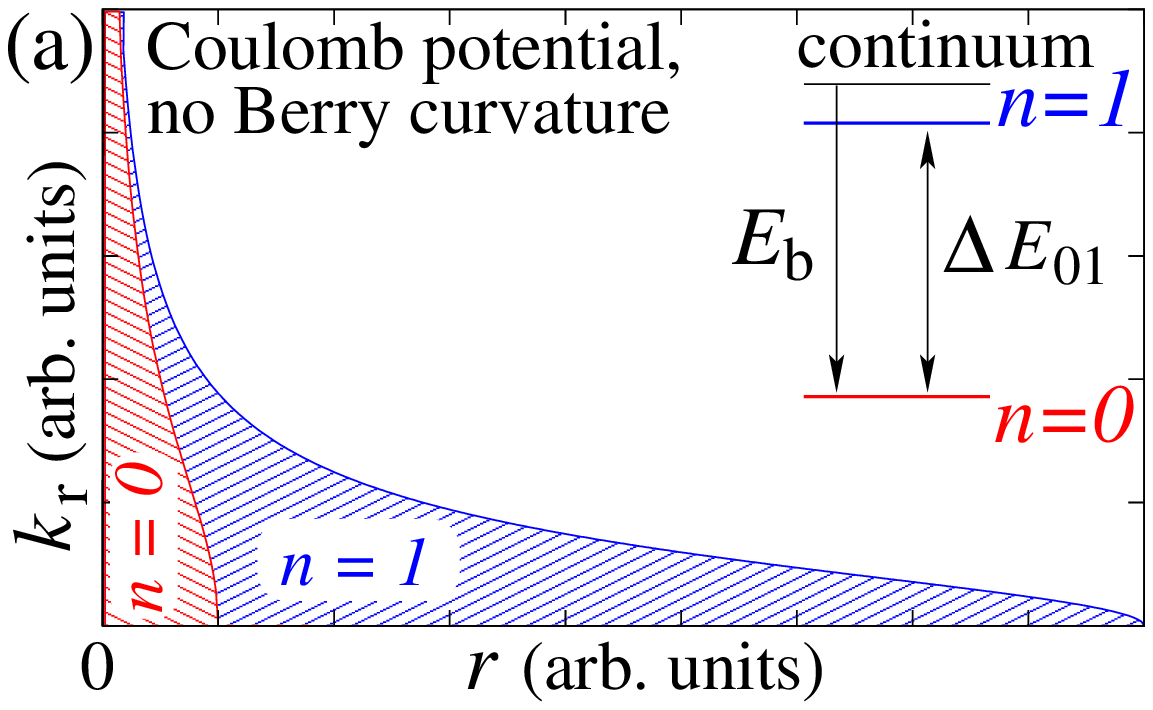}
\includegraphics[width=0.49\columnwidth]{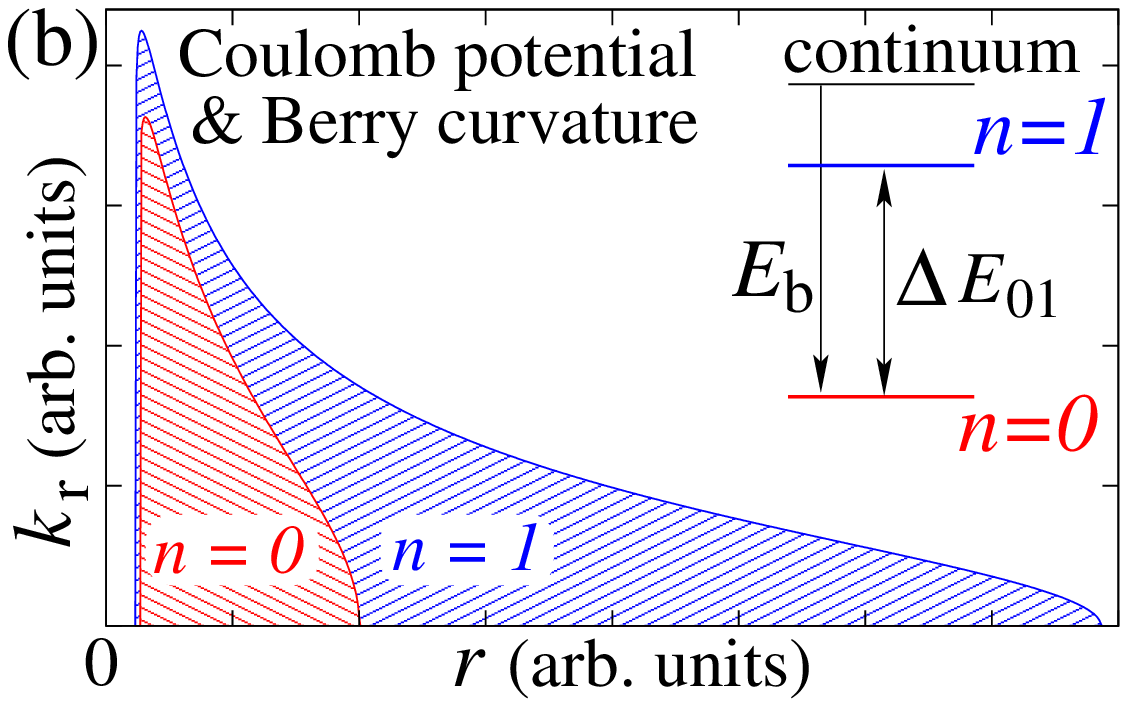}
\includegraphics[width=0.495\columnwidth]{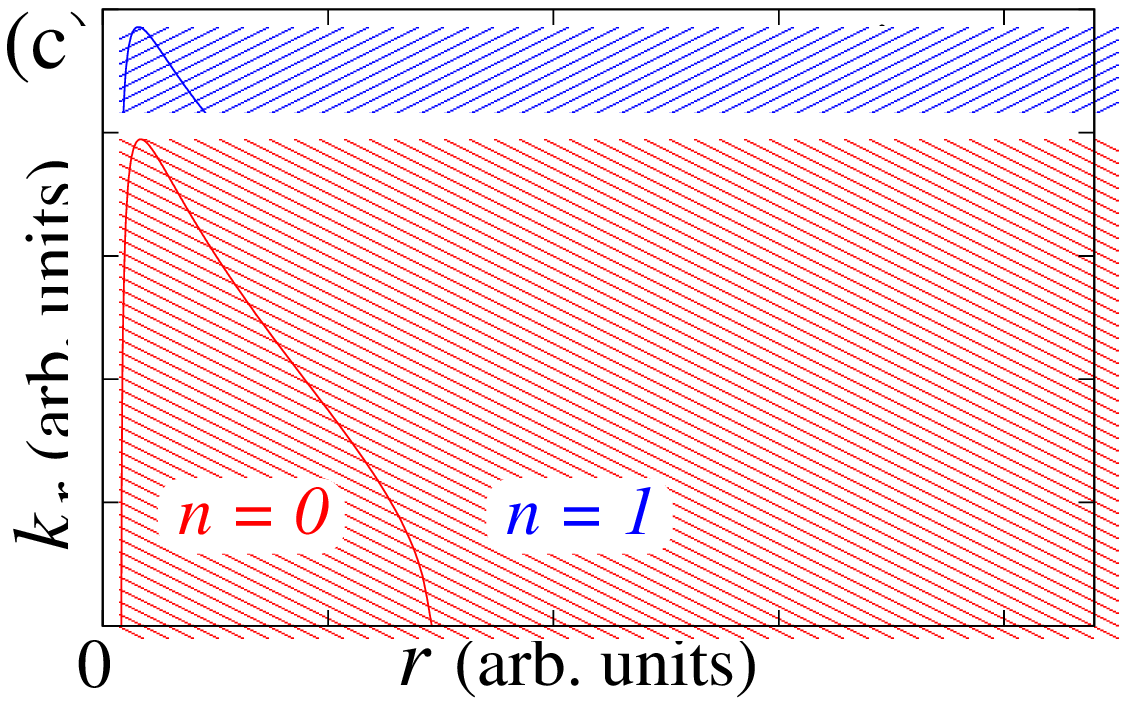}
\includegraphics[width=0.49\columnwidth]{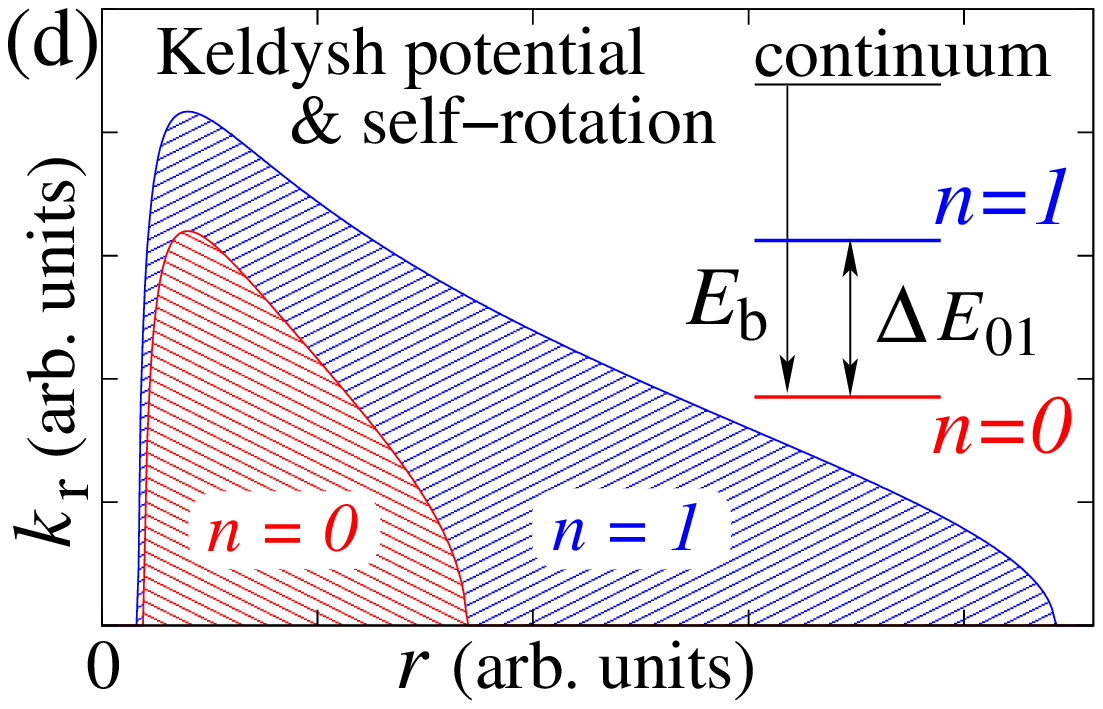}
\caption{\label{fig2} The plots show the phase space available for relative quasiclassical motion of a massive Dirac e-h pair
in the ground and first excited $s$-states as well as the relative energy spacing between them.
The relative spacing $\Delta E_{01}/E_b$ is proportional to the ratio between the two shaded areas. 
Parameters are taken to be relevant for WS$_2$.
(a) Here, the Coulomb interaction with the effective dielectric constant $\epsilon=5.2$ adjusted 
to fit the binding energy measured \cite{PRL2014chernikov}.
Neither the dielectric constant nor the level spacing turn out to be compatible with the values measured.
(b) The Coulomb-Berry model results in a reduced phase space ratio and, as a consequence, reduced relative level spacing
between the $n=0$ and $n=1$ states. The effective dielectric constant taken to fit the binding energy 
is twice smaller than in (a). 
(c) Both Berry curvature and Keldysh potential are taken into account on the same footing.
The relative level spacing (and the phase space ratio) is reduced even stronger, matching the measured spectrum at 
the screening radius of a few nm.
(d) An approximate model where the Berry curvature is emulated by an internal angular momentum $j=1/2$ 
proving the concept of the self-rotating excitons in WS$_2$.}
\end{figure}

{\em Introduction.---} An exciton is a bound electron-hole (e-h) pair optically excited in semiconductors.
In most semiconductors, the electrons and holes behave like conventional Schr\"odinger quasiparticles and
the corresponding exciton energy spectrum represents a hydrogenlike Rydberg series  \cite{Elliott1957}.
In contrast, the electrons and holes in two-dimensional (2D) transition-metal dichalcogenides (TMDs) mimic massive Dirac quasiparticles \cite{PRL2012xiao}
resulting in an intrinsic quantum mechanical entanglement between conduction and valence band states as well as in a finite
Berry curvature entering the quasiclassical equation of motion \cite{PRL2015zhou,PRL2015srivastava,PRB2017gong,SREP2017gong}. 
Moreover, the nonlocal screening of the Coulomb interactions in thin dielectric layers \cite{Keldysh} suggests deviations
from the conventional $1/r$ behavior at small e-h distances $r\sim r_0$, 
where $r_0=2\pi\chi$ is determined by the 2D polarizability $\chi$ \cite{PRB2011cudazzo}.
Here, we show that both ingredients --- the Berry curvature and the
nonlocal screening --- are crucial for a realistic exciton model aimed to describe the exciton energy spectrum in 2DTMDs.
In what follows we employ a quasiclassical approach where the Berry
curvature \cite{PRL2015zhou} and the 2D Keldysh potential \cite{PRB2011cudazzo}
are taken into account on equal footing while quantizing the e-h relative motion. 
We utilize the model to fit the spectroscopic measurements for WS$_2$ \cite{PRL2014chernikov} and  WSe$_2$ \cite{PRL2014he}, see Fig.~\ref{fig1},
employing the only fitting parameter $r_0$. The resulting $r_0\sim 4$
nm agrees with the {\it ab initio} prediction \cite{PRB2013berkelbach}.
If the Berry curvature were neglected, $r_0$ has to be taken twice as large to fit the same spectra.
This discrepancy has been pointed out by Chernikov {\it et al.} in Ref.~\cite{PRL2014chernikov}, where $r_0\sim 8$ nm has been employed.
Our model is able to explain why the polarizability required to fit the measurements \cite{PRL2014chernikov,PRL2014he}
is in fact twice smaller ($\chi\sim 0.7$ nm), in accordance with {\it ab initio} predictions \cite{PRB2013berkelbach}.
The reason is that the Berry curvature results in a centrifugal-like term in the effective quasiclassical Hamiltonian,
hence, reducing the quasiclassical phase space available for the
relative e-h motion {\it as if} there were an additional angular momentum even for $s$-states, see Fig.~\ref{fig2}.
In this sense, the excitons turn out to be self-rotating.
The reduced phase space makes the energy level spacing smaller, similar to what the nonlocal screening (i.e., the Keldysh potential) does.
In what follows we consider the model in detail.

{\em Approach.---} The two-body problem for Dirac particles is not trivial \cite{PRB2010sabio,PRA2013berman,PRB2015berkelbach,PRB2015wu,PRB2014berghauser,JoP2015stroucken,PRB2017demartino}
because of the intrinsic coupling between conduction and valence band
states making the electron and hole states entangled even without Coulomb interactions. 
We employ an effective exciton Hamiltonian that accounts for this coupling 
and has several advantages over that proposed in Ref. \cite{PRB2013rodin}; see the Supplemental Material \cite{SM} for details. 
The Hamiltonian can be written as  $H=H_0+V(r)$, where $V(r)$ is the e-h interaction in relative
coordinates, and $H_0$ is given by 
\begin{equation}
\label{H0}
H_0= \left(\begin{array}{cc}  \frac{\hbar^2 k^2}{2 \mu} & \hbar k \sqrt{\frac{\Delta}{2\mu}} \mathrm{e}^{-i \theta} \\
\hbar k \sqrt{\frac{\Delta}{2\mu}} \mathrm{e}^{i \theta} & \Delta
\end{array}\right).
\end{equation}
Here, $\Delta$ is the band gap, $\mu$ is the exciton reduced mass, and $\tan\left(\theta\right)=k_y/k_x$. 
The spectrum of $H_0$ has two branches:
a single parabolic branch $E_k=\Delta + \frac{\hbar^2 k^2 }{2 \mu}$
describing excitonic states and, in contrast to Ref. \cite{PRB2013rodin},  a dispersionless band $E_0=0$ being the vacuum state from which the excitons are initially excited.
The corresponding eigenstates of $H_0$ can be written as $u_k=(\cos\left(\frac{\Phi}{2}\right), \sin\left(\frac{\Phi}{2}\right) \mathrm{e}^{i\theta})^T$ and
$u_0=(\sin\left(\frac{\Phi}{2}\right), -\cos\left(\frac{\Phi}{2}\right)\mathrm{e}^{i\theta})^T$,
where $\tan\left(\frac{\Phi}{2}\right) = \sqrt{\frac{2\mu\Delta}{\hbar^2 k^2}}$.
In order to make the model analytically tractable even for arbitrary $V(r)$ we rewrite $H$ in the quasiclassical form \cite{PRL2015zhou}
\begin{equation}
 \label{Heff}
 H_\mathrm{exc} = \Delta + \frac{\hbar^2 k^2}{2\mu} + \frac{1}{2}\Omega \cdot (\nabla V \times k) + \frac{1}{4} \Omega \nabla^2 V + V(r), 
\end{equation}
where $\Omega$ is the exciton Berry curvature \cite{PRL2008niu,PRB2011garate}. 
The latter can be computed as \cite{EPL2014goerbig} $\Omega=\nabla_k \times \mathbf{A}$, with  
$\mathbf{A}$ being the Berry connection $\mathbf{A}=i\langle u_k |\nabla_k| u_k \rangle=-\sin^2\left(\frac{\Phi}{2}\right)\nabla_k \theta$.
While the semiclassical expression (\ref{Heff}) is general, the precise form of the quantum Hamiltonian (\ref{H0}) yields
the particular Berry curvature $\Omega_z=\hbar^2\Delta/(\mu E_k^2) \simeq \hbar^2/(\mu\Delta)$ that is twice the one-particle Berry curvature
\cite{PRL2007xiao,PRL2012xiao,EPL2014goerbig}, in agreement with Ref. \cite{PRL2015zhou}.
The two-particle Berry curvature introduced here is a quantity describing the topology 
of an e-h pair, {\em not} of an electron and a hole separately.
Since the potential $V(r)$ is circularly symmetric we employ cylindrical coordinates.
To first order in $\partial_r V$, the total quasiclassical energy then reads
\begin{equation}
\label{Etot}
 E_\mathrm{tot} = E_k   + V(r) + \frac{\hbar^2 \Delta}{2\mu r E_k^2} \frac{\partial V}{\partial r} \left(m + \frac{1}{2}\right),
\end{equation}
where $m=0,\pm 1,\pm 2, ...$ is the magnetic quantum number, and $E_k=\Delta + \frac{\hbar^2 }{2 \mu}\left(k_r^2 + \frac{m^2}{r^2}\right)$.
One notices that the last term is due to the Berry curvature that couples the quantum number $m$ and lifts the $m\leftrightarrow -m$ degeneracy as
pointed out in Refs. \cite{PRL2015zhou,PRL2015srivastava}. The $1/2$ offset is furthermore due to the Darwin-like term $\Omega\nabla^2V/4$ in Eq. (\ref{Heff}), and one obtains
the hydrogen model in the large-gap limit with $\Delta\rightarrow \infty$. 
Solving Eq.~(\ref{Etot}) with respect to the radial wave vector $k_r$, we then employ the Bohr-Sommerfeld quantization rule that in our case results in 
\begin{equation}
 \label{quant}
 \int\limits_{r_1}^{r_2} k_r dr=\pi\left(n + \frac{1}{2} \right),
\end{equation}
where $n=0,1,2, ...$ is the radial quantum number and $r_{1,2}$ are the quasiclassical turning points.
Note that there are in general six solutions for $k_r$ but only one is real and positive.
As we need only $s$-states to compare with experiments, we set $m=0$ everywhere from now on
and consider a set of four models gradually approaching a simple but realistic one.

{\em Coulomb model without Berry curvature.---} This is the simplest exciton model possible, where
the second term in Eq.~(\ref{Etot}) is neglected, and  $V=-e^2/(\epsilon r)$. Equation (\ref{quant}) then explicitly reads 
\begin{equation}
 \label{Coulomb0}
 \int\limits_{r_1}^{r_2} dr \sqrt{\frac{2\mu}{\hbar^2}\left(E_\mathrm{tot} - \Delta  + \frac{e^2}{\epsilon r}  \right)} =\pi\left(n + \frac{1}{2} \right),
\end{equation}
where $r_1=0$, as there is no tangential momentum, see Fig.~\ref{fig2}(a), and $r_2=e^2/\epsilon(\Delta- E_\mathrm{tot})$.
This results in the spectrum 
\begin{equation}
 \label{spectrum0}
 E_n = \Delta - \frac{e^4 \mu}{2\epsilon^2 \hbar^2} \frac{1}{(n+1/2)^2},
\end{equation}
where $n=0,1,2, ...$ is the radial quantum number. In order order to fit the binding energy for WS$_2$ ($0.32$ eV \cite{PRL2014chernikov})
we have to take rather unrealistic effective dielectric constant $\epsilon=5.2$.
Moreover, the level spacing remains heavily overestimated, see
Fig.~\ref{fig1} and Fig.~\ref{fig2}(a). These observations all together 
indicate that the standard exciton model is not suitable for 2DTMDs. 

{\em Coulomb-Berry model.---} Here, we retain the Berry curvature in Eq.~(\ref{Etot}), which for the Coulomb interaction reads
\begin{equation}
 \label{Etot1}
 E_\mathrm{tot} = \Delta + \frac{\hbar^2 k_r^2}{2\mu} +\frac{e^2 \hbar^2 \Delta}{4\epsilon \mu r^3}
 \left( \Delta + \frac{\hbar^2 k_r^2}{2\mu}\right)^{-2} -\frac{e^2}{\epsilon r}.
\end{equation}
$E_\mathrm{tot}$ contains now a centrifugal-like term that makes the phase
space near $r=0$ classically unavailable similar to what the conventional centrifugal term would do once $m\neq 0$.
This is the most important intrinsic property of a self-rotating
exciton that we believe to be crucial for a realistic model. 
Indeed, a real positive solution of Eq.~(\ref{Etot1}) with respect to $\Delta + \frac{\hbar^2 k_r^2}{2\mu}$
(and, hence, with respect to $k_r$) is only possible when
\begin{equation}
\label{cond1}
E_\mathrm{tot} +\frac{e^2}{\epsilon r} \geq \frac{3}{2r}\sqrt[3]{\frac{\hbar^2e^2\Delta}{2\epsilon \mu}}. 
\end{equation}
As consequence, the e-h distance $r$ cannot be smaller than $r_1$, where
\begin{equation}
 \label{rlim}
 r_1 = \frac{1}{E_\mathrm{tot}}\left(\frac{3}{2} \sqrt[3]{\frac{e^2 \hbar^2 \Delta}{2\epsilon \mu}} - \frac{e^2}{\epsilon}\right).
\end{equation}
Since $r_1$ must be positive we find that the effective Bohr radius $r_B=\epsilon \hbar^2/(\mu e^2)$
must be larger than the effective Compton length $\lambda_C =\hbar/\sqrt{\mu\Delta}$.
(The exact condition reads: $r_B > \lambda_C \sqrt{16/27}$.)
The Compton length $\lambda_C$ has a similar meaning here as in high-energy physics:
It constitutes the cutoff below which spontaneous particle creation and annihilation processes become important
and the concept of an exciton as a two-particle system is no longer valid.
From the band theory point of view, this corresponds to the critical
regime when the exciton binding energy approaches the size of the band gap. 
The realistic parameters we employ in Fig.~\ref{fig1} and Table~\ref{tab1} suggest that
both $\lambda_C$ and $r_B$ are a few angstrom that makes our excitons strongly nonhydrogenic even if the interaction is Coulomb-like.

Note that the Berry curvature is hidden not only in the integrand of Eq.~(\ref{quant}) but in its limits as well.
On the one hand, this makes an explicit solution rather cumbersome but possible. Indeed,
Eq.~(\ref{Etot1}) can be solved explicitly with respect to $k_r$ with the relevant branch given by
$$
k_r=\sqrt{\frac{2\mu}{\hbar^2}\left(\frac{a}{3}+ \frac{1+i\sqrt{3}}{6}c + \frac{2}{3}\frac{a^2}{(1+i\sqrt{3})c} -\Delta\right)},
$$
where 
$$
c=\sqrt[3]{\frac{27 b- 2a^3 - \sqrt{(27 b- 2a^3)^2 -4a^6}}{2}},
$$
and $a=E_\mathrm{tot} + e^2/(\epsilon r)$, $b=(e^2 \hbar^2 \Delta)/(4\epsilon \mu r^3)$.
The lower limit is given by Eq.~(\ref{rlim}), whereas
the upper limit can be {\em approximately} taken equal to $r_2$ given below Eq.~(\ref{Coulomb0}).
The latter is possible because the Berry term (being proportional to $1/r^3$) decreases much faster than the Coulomb potential at $r\to \infty$.
On the other hand, the topological centrifugal effect occurs due to the {\em pseudospin}
--- the quantity associated with the $2\times 2$ matrix structure of the effective Hamiltonian (\ref{H0}).
Similar to the real spin the pseudospin can be seen as an internal angular momentum of our exciton.
The explicit diagonalization of our initial Hamiltonian $H$ has been performed 
in Ref.~\cite{PRB2016trushin} within the shallow bound state approximation resulting in the $s$-states spectrum given by 
\begin{equation}
\label{old}
E_n = \Delta - \frac{e^4 \mu}{2\epsilon^2 \hbar^2} \frac{1}{(n+|j|+1/2)^2}, 
\end{equation}
where $j=1/2$ is the pseudospin quantum number. In Table~\ref{tab1}
we compare the exciton energy spectra calculated from Eqs. (\ref{old}) and (\ref{quant})
and find very good agreement for all excited states. The models do not agree on the ground state energy
because of at least two reasons: (i) the quasiclassical approximation is unreliable at low energies,
(ii) the shallow bound state condition $(\Delta-E_n)/\Delta\ll 1$ may not be satisfied for $n=0$.
The latter may indeed take place for WSe$_2$ where the band gap $\Delta=2.02$ eV \cite{PRL2014he} is somewhat 
smaller than in WS$_2$ ($\Delta=2.41$ eV \cite{PRL2014chernikov}) hence resulting
in a stronger discrepancy between the quasiclassical and quantum models.

\begin{table}
\begin{tabular}{llllll}
$|E_n-\Delta|$, eV & $n=0$ & $n=1$ & $n=2$ & $n=3$ & $n=4$\\
\hline
WS$_2$ (self-rot.) & 0.320 & 0.080 & 0.035(5) & 0.020 & 0.0128\\
WS$_2$ (Berry)  & 0.315 & 0.081  & 0.036 & 0.020 &  0.013\\
\hline
WSe$_2$ (self-rot.) & 0.370 & 0.0925 & 0.041(1) & 0.023 & 0.014\\
WSe$_2$ (Berry)  & 0.344 & 0.0922 & 0.041 & 0.023 & 0.014
\end{tabular} 
\caption{\label{tab1} The $j=1/2$ effective self-rotating exciton spectrum (\ref{old})
is a good approximation for the quasiclassical Coulomb-Berry model introduced here. The dielectric constant is
chosen to fit the binding energies $E_b$ measured in Refs. \cite{PRL2014chernikov} and \cite{PRL2014he}:
$\epsilon=2.6$ ($E_b=0.32$ eV, $\Delta=2.41$ eV) for WS$_2$, and $\epsilon=2.5$ ($E_b=0.37$ eV, $\Delta=2.02$ eV) for WSe$_2$.
The exciton reduced mass is the same as in Fig.~\ref{fig1}.}
\end{table}

Figure \ref{fig2}(b) shows how the Berry curvature reshapes the
quasiclassically available phase space for the ground and lowest
excited states. 
As compared to Fig.~\ref{fig2}(a), the ratio between the two substantially decreases, hence,
reducing the relative level spacing. If we again try to fit the binding energy for WS$_2$ by 
tuning the dielectric constant, we obtain the very reasonable number $\epsilon=2.52$ close
to what one expects for graphene on SiO$_2$ substrate
($\epsilon_\mathrm{SiO_2}\sim 4$) with the upper surface exposed to
air ($\epsilon_\mathrm{Air}\sim 1$) 
resulting in $\epsilon\sim 2.5$. 
However, the level spacing is not reduced strong enough to describe
the exciton spectra on a quantitative level. 
To improve the predictive power of our approach we change the
screening model. 

{\em Keldysh-Berry model.---}
Until now we have assumed that the dielectric function is local in real space, i.e.,
$q$ independent in reciprocal space. This is not true for 2D semiconducting systems,
where the dielectric function is given by $\epsilon_q = 1+2\pi \chi q$
\cite{PRB2011cudazzo,PRB2015latini,PRL2016olsen}. 
The real-space e-h interaction potential is known as the Keldysh
potential \cite{Keldysh} given by 
\begin{equation}
 \label{Keldysh}
 V(r) = -\frac{\pi e^2}{2r_0}\left[H_0\left(\frac{r}{r_0}\right) - Y_0\left(\frac{r}{r_0}\right)\right],
\end{equation}
where $H_0$ is the Struve function, $Y_0$ is the Bessel function of the 2nd kind, and $r_0$
is the screening radius being the only fitting parameter. The Berry term in Eq.~(\ref{Etot})
contains now $\partial_r V$ given by
\begin{equation}
 \label{dkeldysh}
 \frac{\partial V}{\partial r} =- \frac{\pi e^2 }{2 r_0^2}\left[Y_1\left(\frac{r}{r_0}\right) + H_{-1}\left(\frac{r}{r_0}\right)\right].
\end{equation}
Because of the complexity of the $r$ dependence in
Eqs.~(\ref{Keldysh}), (\ref{dkeldysh}), it is no longer possible to
obtain explicit expressions for  
 $r_{1,2}$ in Eq.~(\ref{quant}), but we can write the condition for the existence 
of real values of $k_r$:
\begin{equation}
 \label{cond2}
 E_\mathrm{tot} - V(r)\geq \frac{3}{2}\sqrt[3]{\frac{\hbar^2 \Delta}{2\mu r} \frac{\partial V}{\partial r}}.
\end{equation}
One can see that this condition cannot be satisfied near $r=0$, which
again confirms the quasiclassical inaccessibility
of this region as if there were a nonzero tangential momentum. The
available phase space can be qualitatively evaluated 
from Fig.~\ref{fig2}(c). The ratio between phase spaces available for the ground and 1st excited states 
is further reduced making the level spacing even smaller than in the Coulomb-Berry model.
The absolute value of binding energy measured in Refs.~\cite{PRL2014chernikov,PRL2014he} is matched at $r_0$ of a few nm
compatible with {\em ab initio} calculations \cite{PRB2013berkelbach}.

{\em Self-rotating exciton model.---}
The Keldysh-Berry model while being realistic is too cumbersome for an express analysis of the experimental data.
Aiming at a simplified model we employ the correspondence between the
topological self-rotation and pseudospin angular momentum $j=1/2$;
see Table \ref{tab1}. Generalizing this correspondence to any circularly symmetric potential,
the quasiclassical energy for the excitonic $s$-states can be written as
\begin{equation}
\label{Etot3}
 E_\mathrm{tot} = \Delta + \frac{\hbar^2 k_r^2}{2\mu} + \frac{\hbar^2 j^2}{2\mu r^2} + V(r),
\end{equation}
where $j=1/2$ is the the pseudospin angular momentum, and $V(r)$ is given by Eq.~(\ref{Keldysh}). 
The third term then mimics the Berry-curvature contribution and offers a 
good fit in the relevant intermediate-coupling regime, 
when $\alpha=e^2\sqrt{\mu/\Delta}/\hbar\epsilon\sim 1$ (or $\lambda_C$
approaches $r_B$). However, it becomes less accurate when $\Delta$ is drastically increased 
($\alpha \ll 1$) and the Berry curvature vanishes as $1/\Delta$. 
In the opposite limit of decreasing $\Delta$ the self-rotating model
becomes again inapplicable as soon as Eq.~(\ref{cond2}) cannot be satisfied for a certain $r>0$. 
This means that the system experiences an abrupt transition 
to a state that is no longer excitonic and that we do not aim to characterize here.
The fitting model (\ref{Etot3}) does not describe this transition but matches the relevant physics once
the system is {\em already} in a self-rotating regime, see the Supplemental Material \cite{SM} for details.

The quantization condition with the effective $j=1/2$ self-rotation reads
\begin{equation}
 \label{Coulomb2}
 \int\limits_{r_1}^{r_2} dr \sqrt{\frac{2\mu}{\hbar^2}\left[E_\mathrm{tot} - \Delta  - V(r) \right] - \frac{j^2}{r^2}} =\pi\left(n + \frac{1}{2} \right),
\end{equation}
where $r_{1,2}$ are determined by imposing the zero condition on the integrand.
The qualitative justification of this model follows from Fig.~\ref{fig2}(d):
the quasiclassically available phase space appears to be very close to
the one obtained within a somewhat more accurate approach depicted in Fig.~\ref{fig2}(c). 
The major merit of the $j=1/2$ effective model is its transparency in
describing the self-rotation and non-Coulomb potential 
on the equal footing. It can easily be employed to fit the exciton
spectra in Dirac materials \cite{Adv2014wehling} once they will be
measured in future. 

We use Eq.~(\ref{Coulomb2}) to compute the energy for a given $n$; see Fig.~\ref{fig1}.
The model perfectly fits the exciton spectrum measured in WS$_2$
employing $r_0\sim 4$ nm, as predicted {\em ab initio} in
\cite{PRB2013berkelbach}.  
The agreement is less good in the case of WSe$_2$. This
may be due to a certain incompatibility between the experimental data
\cite{PRL2014chernikov,PRL2014he} and 
{\em ab initio} predictions \cite{PRB2013berkelbach}: The exciton
binding energy measured in WSe$_2$ turns out 
to be higher than in WS$_2$, whereas the {\em ab initio} theory
\cite{PRB2013berkelbach} predicts an opposite trend. 
Once the band gap $\Delta=2.02$ eV (and, hence, the binding energy
$E_b=0.37$ eV) is reduced by a few tens of meV  
the exciton spectrum can be fitted for WSe$_2$ as good as for WS$_2$.
However, we emphasize that our model fits the data anyway much better than the conventional exciton model without Berry curvature.

{\em Conclusion.---} 
The very fact that the exciton spectrum in 2DTMDs \cite{PRL2014chernikov,PRL2014he,NanoLett2015hill}
does not resemble the conventional Rydberg series has been discussed
in multiple papers recently
\cite{PRL2016olsen,PRB2015berkelbach,PRB2015wu,PRB2014berghauser,JoP2015stroucken,PRL2015zhou,PRL2015srivastava,trushin2017,Nature2014ye,PRL2013qiu,PRB2012ramasubramaniam,PRB2012komsa,PRB2013shi,PRB2012cheiwchanchamnangij,PRB2016echeverry,2DM2015wang}. The
commonly-accepted explanation \cite{PRL2014chernikov} in terms of the
nonlocal screening  
of the bare Coulomb potential \cite{PRB2011cudazzo} fits well the
exciton spectrum measured \cite{PRL2014chernikov} 
as long as an unrealistically high 2D polarizability is assumed.
To cure this inconsistency, contributions from the Berry curvature need to be considered.
The Berry curvature provides a strong repulsion when $r\rightarrow 0$ even for $s$-states with no angular momentum.
We have shown that this repulsion can be accounted for in a model where
the excitons in 2DTMDs are {\em self-rotating}
in a way similar to the quasiparticles on the surface of a strong topological insulator \cite{PRB2011garate},
thus opening further perspectives for excitonic engineering in van der
Waals heterostructures \cite{PRL2017wu}.

We acknowledge financial support from the Center of Applied Photonics funded by the DFG-Excellence Initiative.
M.T. was supported by the Director's Senior Research Fellowship from the Centre for Advanced 2D Materials at the National University of Singapore
(NRF Medium Sized Centre Programme R-723-000-001-281).
We thank Alexey Chernikov and Oleg Berman for discussions.

\bibliography{excitons-theory.bib,excitons-experiment.bib}

\appendix

\section{Effective exciton Hamiltonian}

Here, we justify the effective excitonic Hamiltonian suggested in the main text.
We focus on the $K$-valley, where free electrons (e) are described by the two-dimensional Dirac Hamiltonian given by
\begin{equation}\label{ham:el}
 H_e=\left(\begin{array}{cc}
M &   \hbar v k e^{-i \theta } \\
 \hbar v k e^{i \theta }  &  -M  \\
\end{array}
\right),\quad 
\end{equation}
where $\tan \theta = k_y/k_x$, and $\Delta=2M$ is the fundamental band gap.
The corresponding free hole (h) Hamiltonian $H_h$ at zero center-of-mass momentum reads \cite{PRB2013rodin}
\begin{equation}\label{ham:h}
 H_h=\left(\begin{array}{cc}
-M &   \hbar v k e^{i \theta } \\
 \hbar v k e^{-i \theta }  &  M  \\
\end{array}
\right).
\end{equation}
Here, we have assumed that the electron and hole have opposite momenta for optical excitons we consider.
The angle $\theta$ has therefore been changed to $\theta+\pi$ for holes. This has been done in \cite{PRB2013rodin} at a later stage.

These two Hamiltonians suggest the same quasiparticle dispersion 
\begin{equation}
\varepsilon_k=\sqrt{(\hbar v k)^2+M^2},
\label{eps_k}
\end{equation}
that in turn suggests the same effective mass $m^*=M/v^2=\Delta/(2v^2)$ for both e and h. 
The kinetic term for the minimal exciton model
with zero center-of-mass momentum should therefore read as $\hbar^2 k^2/(2\mu)$,
where $\mu=m_e m_h /(m_e+m_h)=m^*/2=M/(2v^2)=\Delta/(4v^2)$ is the reduced effective mass.
This model, however, does not take into account the pseudospin degree of freedom
that makes the electron and hole states entangled even without Coulomb interaction.

The correct two-particle e-h Hamiltonian with vanishing interaction is given by the Kronecker sum
$H_4=H_e \otimes I_2 + I_2 \otimes H_h $ with $I_2$ being the $2\times 2$  unit matrix.
Since the one-particle electron Hamiltonian $H_e$ already has an electron and a hole branch, one might not see the "need" to
extend the dimension of our Hilbert space. However, this is a flaw since electron and hole branches
are intimately related on the single-particle level in Dirac-like Hamiltonians such that they do not
represent two different particles. That is why the two-body problem consists of a tensor product of
two Dirac particles, each of which has its own electron and hole branch.
Hence, we have \cite{PRB2013rodin}
\begin{equation}
H_4=\left( \begin{array}{cccc}
 0 & e^{i \theta } \hbar k v & e^{-i \theta } \hbar k v & 0 \\
 e^{-i \theta } \hbar k v & 2 M & 0 & e^{-i \theta } \hbar k v \\
 e^{i \theta } \hbar k v & 0 & -2 M & e^{i \theta } \hbar k v \\
 0 & e^{i \theta } \hbar k v & e^{-i \theta } \hbar k v & 0 \\
\end{array}
\right).
\end{equation}
The Hamiltonian $H_4$ has four eigenvalues: 
$$\left\{0,0,+2\sqrt{\hbar^2 v^2 k^2 +M^2},-2\sqrt{\hbar^2 v^2 k^2 +M^2}\right\}.$$
The physical meaning of the four states can be analyzed having two excited
particles in play. The exciton state corresponds to an electron in the conduction band and a hole in
the valence band. One of two zero-energy states corresponds to the exciton vacuum when all the
holes are in the conduction band and all the electrons are in the valence band. The second zero-energy
state corresponds to the inverse population (inverted vacuum) when all the electrons are in the conduction
band and all the holes are in the valence band. Finally, the negative-dispersion branch corresponds
to a single electron in the valence band and a single hole in the conduction band that is an exciton
created in the inverted vacuum. Obviously, the latter two configurations are not relevant for our
setting with the low-intensity optical excitations.
Thus, we extract the excitonic part out of the Hamiltonian $H_4$. This is done in two steps. First, we extract the relevant $2\times 2$ block via an
appropriate unitary transformation. This block contains the positive excitonic band plus a zero-energy flat band. While we are interested, from 
a purely spectral point of view, only in the positive branch, the coupling to the zero-energy band bears relevant information about the Berry-curvature
contributions that we discuss in more detail in the following section. 
In a second step, we consider the $2\times 2$ block up to second order in momentum. 
Its Dirac-type structure allows us to calculate directly the Berry curvature of the exciton branch. The consistency of its expression with that of 
a quasiclassical approach corroborates the validity of the effective quantum Hamiltonian obtained within our work.

In detail, to transform $H_4$ into the block-diagonal form, we use the transformation matrix
\begin{widetext}
\begin{equation}
P=\left( \begin{array}{cccc}
 \cos\left(\frac{\Theta}{2}\right) & 0 & \sin\left(\frac{\Theta}{2}\right) & 0 \\
 0 & \cos\left(\frac{\Theta}{2}\right) & 0 & \sin\left(\frac{\Theta}{2}\right) \\
 e^{i \theta } \sin\left(\frac{\Theta}{2}\right) & 0 & -e^{i \theta } \cos\left(\frac{\Theta}{2}\right) & 0 \\
 0 & e^{i \theta } \sin\left(\frac{\Theta}{2}\right) & 0 & -e^{i \theta} \cos\left(\frac{\Theta}{2}\right) \\
\end{array} \right),
\end{equation}
\end{widetext}
where
$$
\tan\Theta=\frac{\hbar v k}{M}.
$$
The transformed matrix $H_4'=P^{-1}H_4 P$ then reads
\begin{equation}
H_4'=  \left( \begin{array}{cccc}
-M+\varepsilon_k  & e^{i \theta } \hbar k v & 0 & 0 \\
 e^{-i \theta } \hbar k v &  M+\varepsilon_k & 0 & 0 \\
  0 & 0 & -M-\varepsilon_k & e^{i \theta } \hbar k v \\
 0 & 0 & e^{-i \theta } \hbar k v & M-\varepsilon_k \\
\end{array}
\right),\\
\end{equation}
where $\varepsilon_k$ is given by equation (\ref{eps_k}).
The upper left block $2\times 2$ has two eigenvalues $\left\{0,+2\sqrt{\hbar^2 v^2 k^2 +M^2}\right\}$ 
describing excitonic quasiparticles we are interested in. The block is decoupled from the rest and can be seen as a reduced excitonic Hamiltonian:
\begin{equation}
H_2 = 
\left(\begin{array}{cc}
-M+\sqrt{\hbar^2 v^2 k^2 +M^2} &  e^{i \theta } \hbar k v \\
 e^{-i \theta } \hbar k v &  M+\sqrt{\hbar^2 v^2 k^2 +M^2}  \\
\end{array}
\right).
\label{correct}
\end{equation}
Note that the dispersion $2\sqrt{\hbar^2 v^2 k^2 +M^2}$ once written in the effective mass approximation suggests
correct reduced mass $\mu=M/(2v^2) =\Delta/(4v^2)$. 
Once the Schr\"odinger terms are neglected and the missing momentum is compensated by the factor of 2 in the off-diagonal terms
we obtain the reduced excitonic Hamiltonian given by
\begin{equation}
H_2^\mathrm{RCN} = 
\left(\begin{array}{cc}
M &  2 \hbar k v e^{i \theta }  \\
 2 \hbar k v e^{-i \theta }  &  -M  \\
\end{array}
\right).
\end{equation}
This form has been suggested by Rodin and Castro Neto in Ref. \cite{PRB2013rodin} but it has two major drawbacks:
\begin{itemize}
 \item Its spectrum $\pm\sqrt{(2\hbar v k)^2 + M^2}$ suggests the effective reduced mass $M/4v^2$ twice smaller than needed ($M/2v^2$). As a consequence,
 the excitonic Berry curvature would read $\Omega^\mathrm{RCN}_\mathrm{exc}(k=0)=2\hbar^2v^2/M^2=8\hbar^2v^2/\Delta^2$, 
 which is twice as large as the one obtained by Zhou {\sl et al.} \cite{PRL2015zhou}.
 \item The lower dispersionless branch describing zero-energy states is substituted by an ``antiparticle'' branch that
 contradicts to physical meaning of the vacuum state from which the excitons are excited.
\end{itemize}

To remedy the situation we must retain both Dirac and Schr\"odinger terms in equation (\ref{correct}).
At the same time, the Hamiltonian must be parametrized in such a way that the lower spectral branch always remains dispersionless. 
The dispersionless vacuum state is of utmost importance for optically excited excitons where
electron and hole momenta are antiparallel and, hence, cancel each other once an exciton recombines.
Any vacuum state with dispersion suggests some uncompensated momentum and violates momentum conservation for direct-gap optical excitons.

We aim for an exciton model in terms of the effective mass, so that we consider the case of small $k$ 
and expand the energy of e-h relative motion up to the terms quadratic in $k$.
If we keep quadratic terms in both lines of the matrix (\ref{correct}), then we end up with the spectrum quartic in $k$.
To avoid this inconvenience we expand each element of the matrix up to the lowest non-zero order as follows:
$$
-M+\sqrt{\hbar^2 v^2 k^2 +M^2}\propto k^2, \quad M+\sqrt{\hbar^2 v^2 k^2 +M^2}  \to 2M.
$$
Then, we modify the Dirac part in order to compensate parasitic terms in the dispersionless branch arising due to the different approximations
done in these two terms, and $H_2$ reads
\begin{equation}
H_2^\mathrm{eff} =
\left(\begin{array}{cc}
\frac{\hbar^2 v^2 k^2}{M} &  e^{i \theta } \sqrt{2}\hbar v  k  \\
 e^{-i \theta } \sqrt{2}\hbar v k  & 2M  \\
\end{array}
\right).
\label{correct2}
\end{equation}
Despite different approximations done in lower and upper lines of equation (\ref{correct2})
the Hamiltonian obtained satisfies all physical criteria we demand:
\begin{itemize}
\item $H_2^\mathrm{eff}$ results in two spectral branches, $E_0=0$ and $E_k= 2M+ \frac{\hbar^2 v^2 k^2}{M}$,
where the latter suggests correct reduced mass, $\mu=M/2v^2$.
\item The vacuum state remains  dispersionless, as it is expected from the correct excitonic Hamiltonian  (\ref{correct})
and physical constrains explained above.
\item The two spectral branches remain entangled, as it is again expected for Dirac excitons.
\item The Hamiltonian contains the Schr\"odinger term that preserves the exciton from a collapse.
\end{itemize}

Equation (\ref{correct2}) can be rewritten in terms of the reduced effective mass $\mu=M/2v^2$ as
\begin{equation}\label{ham:eff}
H_2^\mathrm{eff} = 
\left(\begin{array}{cc}
\frac{\hbar^2 k^2}{2\mu} &  e^{i \theta } \hbar k \sqrt{\frac{M}{\mu}} \\
 e^{-i \theta } \hbar k \sqrt{\frac{M}{\mu}} & 2M  \\
\end{array}
\right).
\end{equation}
We use this version in our quasiclassical model as an effective exciton Hamiltonian with vanishing interaction.
It is clear that, once $\mu$ is fixed, there is only one possible form of $H_2^\mathrm{eff}$
providing a dispersionless branch for the vacuum state. 
One can easily obtain the Berry curvature of the exciton band, which is written at $k=0$ as
\begin{equation}
\label{Omegaexc}
 \Omega_\mathrm{exc}^{\mathrm{eff}}(k=0)= \frac{\hbar^2}{\mu\Delta}=2\frac{\hbar^2}{m^*\Delta}=2\Omega_e(k=0),
\end{equation}
i.e. it is twice the electronic Berry curvature $\Omega_e(k=0)=\hbar^2/m^*\Delta$ derived form $H_e$, see equation (\ref{ham:el}). This agrees with the quasiclassical 
analysis of Ref. \cite{PRL2015zhou}, where the exciton Berry curvature was shown to be $\Omega_\mathrm{exc}(k=0)=\Omega_e(k=0) - \Omega_h(k=0)$,
in terms of the Berry curvature of the one-particle Dirac models (\ref{ham:el}) and (\ref{ham:h}), for the electron and the hole, respectively.
Since in this case, one has $\Omega_e(k)= - \Omega_{h}(k)$, one obtains $\Omega_\mathrm{exc}(k=0)=2\Omega_e(k=0)$, in agreement 
with the Berry curvature obtained from our effective $2\times 2$ Hamiltonian (\ref{ham:eff}).

\section{Semi-quantitative justification for the self-rotating exciton model in the intermediate coupling regime}

We now use the semi-classical exciton Hamiltonian derived in Ref. \cite{PRL2015zhou} taking into account the Berry curvature,
\begin{eqnarray}\label{eq:hamB}
  {H}_\mathrm{exc} &=& \Delta + \frac{\bp}{2\mu} +  {V}(r) \\
 \nonumber
 &&+ \frac{1}{2\hbar}  \Omega(\bp)\cdot \left[\frac{\partial  {V}(r)}{\partial \mathbf{r}}\times \bp\right]
 +\frac{1}{4}\Omega(\bp)\nabla^2 {V}(r),
\end{eqnarray}
in terms of the relative coordinate $\mathbf{r}$ and momentum $\bp=\hbar \bk$ of the exciton, 
and use 
\begin{equation}
 \Omega_z(\bp=0)\equiv\Omega_\mathrm{exc}^{\mathrm{eff}}(k=0)
\end{equation}
for the Berry curvature, see equation (\ref{Omegaexc}). We furthermore introduce the ``Compton length'' 
$\lambda_C=\hbar/\sqrt{\Delta\mu}$, which constitutes one of the natural length scales of the problem (it could also be called the gap length here)
and that we will use in the expression of the Hamiltonian, which reads (in cylinder coordinates)
\begin{eqnarray}
\nonumber
 {H}_\mathrm{exc} &=& \Delta\left[1 + \frac{\lambda_C^2}{2}\left( {k}_r^2 + \frac{m^2}{r^2}\right)\right] \\
&&+  {V}(r) + \frac{\lambda_C^2}{2r}\frac{\partial  {V}}{\partial r}
 \left(m+\frac{1}{2}\right),
\end{eqnarray}
where $m=0,\pm 1,\pm 2,...$ is the quantum number of the tangential momentum, and
$k_r$ is the radial wave-vector. If we now consider a Coulomb interaction $ {V}(r)= -e^2/\epsilon r=- \alpha\Delta(\lambda_C/r)$,
in terms of the dimensionless coupling strength $\alpha=(e^2/\hbar\epsilon)\sqrt{\mu/\Delta}$, one can write down the Hamiltonian in a dimensionless
manner
\begin{eqnarray}
\nonumber\label{eq:HamB0}
 \frac{ {H}_\mathrm{exc}}{\Delta} &=& 1+ \frac{1}{2}\left(\lambda_C^2k_r^2 + m^2 \frac{\lambda_C^2}{r^2}\right) \\
 &&-\alpha \frac{\lambda_C}{r} +\frac{\alpha}{2}\left(m+\frac{1}{2}\right)\left(\frac{\lambda_C}{r}\right)^3.
\end{eqnarray}
Notice that one could also have chosen the Bohr radius $r_B=\hbar^2\epsilon/\mu e^2$ as a natural length scale. This choice is mostly made in the context of the 2D 
hydrogen model. It is related to the Compton length by $\lambda_C = \alpha r_B$ -- a third length scale is given by the comparison between interaction and gap, 
$\ell=e^2/\epsilon\Delta$, and one finds the hierarchy
\begin{equation}
 \ell = \alpha \lambda_C=\alpha^2 r_B.
\end{equation}
The Compton length has a similar meaning here as in high-energy physics. It constitutes the lower bound of the length scale above which the exciton problem can be treated in 
terms of relativistic quantum mechanics, such as in our present approach. On length scales lower than $\lambda_C$, one needs to take into account spontaneous electron-hole creation
processes that could eventually provide (on the average) additional screening to the effective interaction potential $ {V}(r)$.
For $\alpha\lesssim 1$, the Bohr radius is indeed larger than $\lambda_C$, and the above-mentioned processes can be safely neglected. This argument in terms of length scales is 
in line with one obtained from energy scales: in order to be allowed to restrict the exciton dynamics to a single band (i.e. an electron in the conduction and a hole in the 
valence band), one needs to ensure that the exciton energy $\sim e^2/\epsilon r_B$ (for a typical exciton size given by the Bohr radius, as for the $s$ state) is small as compared 
to the gap $\Delta$, but the ratio is precisely given by $e^2/\epsilon r_B\Delta=\ell/r_B=\alpha^2$, such that one obtains again $\alpha\lesssim 1$. 

While the above expression (\ref{eq:HamB0})
is well adapted to connect the problem to the non-interacting case, by setting $\alpha\rightarrow 0$, it does not really allow us to appreciate the 
relative role of the two interaction terms. In units of $r_B$, the Hamiltonian then reads 
\begin{eqnarray}
\nonumber
 \frac{{H}_\mathrm{exc}}{\Delta} &=& 1+ \frac{\alpha^2}{2}\left(r_B^2k_r^2 + m^2 \frac{r_B^2}{r^2}\right) \\
 &&-\alpha^2 \frac{r_B}{r} +\frac{\alpha^4}{2}\left(m+\frac{1}{2}\right)\left(\frac{r_B}{r}\right)^3,
\end{eqnarray}
and one now realizes that in the weak-coupling limit ($\alpha\ll 1$), where the typical length scale is indeed set by the Bohr radius, one retrieves the 2D hydrogen problem since
the Berry-curvature term becomes less relevant ($\alpha^4$ instead of the leading $\alpha^2$). This may be seen as a consistency test of the quasiclassical treatment 
since one expects indeed to retrieve the hydrogen problem, described in terms of a simple one-band Schr\"odinger equation, in this limit.
In our case (2D transition-metal dichalcogenides), however, we are in
an intermediate-coupling regime, with $\Delta\sim 2$ eV and $m_e\simeq m_h\simeq 0.1 m_0$
($m_0$ is the bare electron mass) suggesting $\alpha\simeq (2...3)/\epsilon$, that is on the order of unity for a typical dielectric
constant of $\epsilon \simeq 2.5$. 

From the above discussion we see that the self-rotating model is a good description in this limit of \textit{intermediate couplings}. Naturally, the Berry-curvature term scales as $(r_B/r)^3$, thus 
prohibiting \textit{strictu sensu} a description in terms of a centrifugal term that scales as $(r_B/r)^2$. However, the typical exciton radius in the $s$ (and even $p$) states
is on the order of $r_B$, for principal quantum numbers that are not too large. In this case, one might try to merge the Berry-curvature and the centrifugal term in a rather crude 
manner,
\begin{equation}
 m^2 + \alpha^2 m + \alpha^2/2 \simeq j^2 + c,
\end{equation}
where $j=(m+\alpha^2/2)$ and we have introduced a corrective term (the error one makes in absorbing the Berry-curvature term in a $j^2$), $c=\alpha^2/2-\alpha^4/4$. While this
corrective term vanishes for $\alpha = \sqrt{2}$, it is $1/4$ for $\alpha=1$ that yields $j=m+1/2$ as stipulated within the self-rotating model. 

While our analysis shows  that the self-rotating approximation is indeed a good one for values of $\alpha\sim 1$, it breaks down in the weak-coupling 
limit, where one retrieves the hydrogen problem. Furthermore, it might also be inappropriate in the discussion of large values of $m$ (and $n$) where the effective radius (i.e.
the extension of the wave function) is substantially larger than $r_B$. From a theoretical point of view, it might therefore be reasonable to compare the different models (just Coulomb for
the moment) in these large-$m$ and large-$n$ limits even if this limit does not seem relevant in the physical properties of excitons in 2D transition-metal dichalcogenides 
on the experimental side. This
would allow us to test the limits of the self-rotating approximation for $\alpha\sim 1$.

\end{document}